\documentclass[nofootinbib, a4paper, aps, pra, reprint]{revtex4-1}

\usepackage[T1]{fontenc}
\usepackage{amsfonts, amsmath, amssymb, amsthm} 
\usepackage{dsfont, setspace, braket, hyperref, multirow, xcolor, enumerate} 
\usepackage{graphics, graphicx, float, bm, lipsum, soul, silence}

\DeclareMathOperator{\tr}{tr}

\makeatletter
\@ifdefinable\CmdName{
    \long\def\CmdName#1#{\romannumeral0\innerCmdName{#1}}
}
\newcommand\innerCmdName[2]{
    \expandafter\exchange\expandafter{\csname #2\endcsname}{ #1}
}
\newcommand\exchange[2]{#2 #1}
\newcommand\definechars[1]{
    \ifx\relax#1\expandafter\@gobble\else\expandafter\@firstofone\fi{
        \CmdName\newcommand*{#1C}{\mathcal{#1}}
        \definechars
    }
}
\makeatother
\definechars ABCDEFGHIJKLMNOPQRSTUVWXYZ\relax

\makeatletter
\renewcommand\ket[1]{
    \@ifnextchar\bra{\k@t{#1}\!}{\k@t{#1}}
    }
\newcommand\k@t[1]{{|{#1}\rangle}}
\makeatother

\hypersetup{colorlinks,
    linkcolor = {blue},
    citecolor = {blue!50!black},
    urlcolor = {blue!80!black},
}

\begin{document}

\title{Machine learning meets the CHSH scenario}

\author{Gabriel Pereira Alves\textsuperscript{1}}
\email{gpereira@fuw.edu.pl}
\author{Nicolas Gigena\textsuperscript{2}}
\email{nicolas.gigena@fisica.unlp.edu.ar}
\author{J\k{e}drzej Kaniewski\textsuperscript{1}}
\email{jkaniewski@fuw.edu.pl}

\affiliation{$^1$Faculty of Physics, University of Warsaw, Pasteura 5, 02-093 Warsaw, Poland}
\affiliation{$^2$IFLP/CONICET and Departamento de F\'{\i}sica, Universidad Nacional de La Plata, C.C. 67, La Plata (1900), Argentina.}

\date{\today}

\begin{abstract}
   In this work, we perform a comprehensive study of the machine learning (ML) methods for the purpose of characterising the quantum set of correlations. As our main focus is on assessing the usefulness and effectiveness of the ML approach, we focus exclusively on the CHSH scenario, both the 4-dimensional variant, for which an analytical solution is known, and the 8-dimensional variant, for which no analytical solution is known, but numerical approaches are relatively well understood. We consider a wide selection of approaches, ranging from simple data science models to dense neural networks. The two classes of models that perform well are support vector machines and dense neural networks, and they are the main focus of this work. We conclude that while it is relatively easy to achieve good performance on average, it is hard to train a model that performs well on the "hard" cases, i.e., points in the vicinity of the boundary of the quantum set. Sadly, these are precisely the cases which are interesting from the academic point of view. In order to improve performance on hard cases one must, especially for the 8-dimensional problem, resort to a tailored choice of training data, which means that we are implicitly feeding our intuition and biases into the model. We feel that this is an important and often overlooked aspect of applying ML models to academic problems, where data generation or data selection is performed according to some implicit subjective criteria. In this way, it is possible to unconsciously steer our model, so that it exhibits features that we are interested in seeing. Hence, special care must be taken while determining whether ML methods can be considered objective and unbiased in the context of academic problems.
\end{abstract}

\maketitle

\section{Introduction}

\noindent Proposed in 1964 \cite{Bell64}, Bell non-locality \cite{Brunner14} establishes that the predictions of quantum mechanics are inconsistent with local hidden-variables (LHV) models, constituting a fundamental aspect of quantum correlations in spatially separated systems. The so-called Clauser-Horne-Shimony-Holt (CHSH) scenario \cite{Clauser69} is the simplest scenario in which the correlations observed by two parties can be verified as not admitting a local-realistic explanation, assuming that the shared quantum state is entangled and that the pair of dichotomic measurements implemented by each party is incompatible \cite{Fine82}. Such correlations are identifiable through the violation of Bell inequalities, and assessing the extent of these violations has always been a question of broad interest. Tsirelson was one of the first to look into the topic, deriving, in 1980 \cite{Tsirelson80}, an upper bound for the violation of the CHSH inequality, a limit that would later come to be known as Tsirelson's bound. In the same direction, Tsirelson himself, in 1987 \cite{Tsirelson87}, followed by others \cite{Landau88, Uffink02, Masanes03}, obtained a tighter characterisation in the form of a nonlinear inequality restricting the set of quantum correlations, currently known as the TLM inequality. The latter, even if satisfied by any quantum correlation, is sufficient only when the local outcomes of Alice and Bob have uniform distributions.

Exploiting the fact that the set of quantum correlations is convex, two complementary heuristic approaches are often used to find maximum values of a given Bell inequality. On the one hand, outer approximations to the quantum set can be obtained through the NPA hierarchy \cite{Navascues07} of semi-definite programs (SDPs), which is guaranteed to converge to the quantum set. In the CHSH scenario with uniformly distributed outcomes, the TLM inequalities can be recovered at the first level of this hierarchy, but it is currently unknown if it converges to the quantum set at a finite level in any other scenario. On the other hand, the see-saw optimisation \cite{Werner01} provides an inner approximation that can also be cast as an SDP \cite{Liang07}. In this latter procedure, from a random initialisation, state and measurements are optimised separately and iteratively, so that after a certain number of iterations the objective function converges to a local maximum. Unlike the NPA approach, the see-saw optimisation depends on the quantum state dimension, and due to its non-deterministic initialisation, the yielded output is also non-deterministic. Both techniques, NPA and see-saw, when applied to a given Bell functional, often converge to the same value \cite{Brunner14}, making this combination a valuable tool for exploring the boundary of the set of quantum correlations.

While the approaches mentioned above were specifically designed for the problem of quantum correlations, machine learning (ML) methods are based on the idea that by providing a sufficient amount of data, it is possible to approach a given problem without relying on any prior intuition. In fact, with the emergence of such tools, many areas of physics \cite{Carleo19} and, more specifically, quantum information \cite{Dunjko18, Bharti20}, have benefited from their application in addressing problems without analytical solutions or even those the computation of which is costly. With Bell non-locality, the use of ML was no different, and advances in quantifying non-locality in Bell scenarios \cite{Canabarro19}, networks \cite{Krivachy20}, or even to assist the solution of feasibility SDPs \cite{Krivachy21} were made, to mention a few.

Building upon the foundational work in non-locality and the advances in ML, this work aims to use ML techniques, specifically support vector machines (SVMs) and neural networks (NNs), to delve deeper into the exploration of the quantum set. Differently from previous methodologies that applied varied methods to specific problems, here we concentrate on the simplest Bell scenario, assessing distinct models to identify the most effective approaches. Our methodology involves using the subspace of the CHSH scenario where the TLM inequalities hold as a benchmark for exploring the entire quantum set, framing the task as a classification problem. As previously pointed out, this region of the quantum set has an analytical description, allowing us to develop a model to be compared with the results already known. By doing so, we aim to achieve an understanding of the models' efficacy in this confined scenario, subsequently extrapolating the trained models to the entire quantum set. This strategy does not only ensure a solid foundation for the models, thanks to the available analytical description, but also allows us to estimate the limits of our techniques in unexplored regions of the quantum set. In essence, we aim to find out which ML methods work best for quantum correlations, what kind of accuracy can be expected, and how expensive they are compared to the currently used methods.

This work is organized as follows. In Sec.~\ref{sec: preliminaries}, we provide a preliminary introduction to the main concepts and notations regarding Bell non-locality, as well as present the basic notions of machine learning. Sec.~\ref{sec: data generation} discusses the data generation methods for correlations only and also for the entire scenario. We conclude this section by commenting on a novel technique to close the gap between NPA and see-saw for non-exposed points in the boundary of the quantum set. Sec. \ref{sec: fitting the model} discusses the ML models utilized, and the results obtained for both SVMs and NNs. We close our considerations in Sec.~\ref{sec: discussion}.

\section{Preliminaries} \label{sec: preliminaries}

\noindent Contrary to what is known for the set of non-signalling ($\NC\SC$) and local ($\LC$) correlations, which in every single-sourced Bell scenario are polytopes, the set of quantum correlations ($\QC$) has a more complex shape. Characterising its boundary is difficult, and a partial solution is known only in the CHSH scenario. In this section, we introduce the established concepts and notations regarding the CHSH scenario, followed by basics of ML.

\subsection{Correlation sets in the CHSH scenario}

The CHSH scenario consists of a bipartite setup where the two parties, Alice and Bob, share a state in which they are allowed to perform two dichotomic measurements. We label Alice's and Bob's measurements by $x, y\in \{0, 1\}$, whereas their respective outcomes by $a, b \in \{ \pm 1\}$. Given the measurements $x$ and $y$, the probability of obtaining outcomes $a$ and $b$, respectively, is written as $p (a, b \,|\, x, y)$. The 16-tuple of all joint probabilities is named \textit{correlation} or \textit{behaviour}, which we can think of as a point in $\mathbb{R}^{16}$. Because of normalisation and non-signalling constraints, the non-signalling polytope lives in an $8$-dimensional subspace of $\mathbb{R}^{16}$, thus making it possible to represent it with a lower dimensional parametrisation. One of the most widely used is that in terms of {\it marginals} and {\it correlators} $\{\braket{A_x}, \braket{B_y}, \braket{A_xB_y}\}_{x,y}$, related to the probabilities via the following expressions:
\begin{equation}
\begin{split}
\braket{A_x B_y} &:= \sum_{a, b \in \{\pm 1\}} \, ab\, p(a, b \,|\, x, y),\\
\braket{A_x} &:= \sum_{a \in \{\pm 1\}} \, a\, p_A(a \,|\, x)\;\;\text{and}\\
\braket{B_y} &:= \sum_{b \in \{\pm 1\}} \, b\, p_B(b \,|\, y),
\label{eq: correlators representation}
\end{split}
\end{equation}
where $p_A(\cdot)$ and $p_B(\cdot)$ denote the marginal distributions of $p(a, b|x, y)$ over $a$ and $b$, respectively. In this case, the set of non-signalling behaviours is represented by a subset of the hypercube in $\mathbb{R}^8$ defined by the coordinates $\braket{A_x}, \braket{B_y}, \braket{A_xB_y} \in [-1, 1]$. Of particular interest for us is the subspace satisfying $\braket{A_x} = \braket{B_y} = 0$, i.e., the restriction of the set of behaviours to the {\it correlation space}, since for this subset the TLM inequalities and the positivity facets fully describe the boundary of the quantum set.

Quantum behaviours are those for which a bipartite state $\rho_{AB}$ and local measurement operators $\{ M_{a|x}, N_{b|y}\}$ exist such that
\begin{equation}
    p(a, b \,|\, x, y) = \tr \big( \rho_{AB} \, M_{a|x} \otimes N_{b|y} \big),\quad \forall\; a, b, x, y.
    \label{quantum_prob}
\end{equation}
We can directly compute the correlators as the expected values of the observables, defined by $A_x = M_{1|x} - M_{-1|x}$ and $B_y = N_{1|y}- N_{-1|y}$, that is,
\begin{equation}
    \braket{A_xB_y}= \tr \big(\rho_{AB} \, A_x \otimes B_y \big).
    \label{eq: correlators}
\end{equation}
We refer to the triple $\{A_x, B_y, \rho_{AB}\}$ as a {\it realisation} and $\QC$ is defined as the set of behaviours for which such a realisation exists.

As already mentioned, the boundary of $\QC$ restricted to the correlation space admits an analytical description in terms of the TLM inequalities, the elements of which can be obtained by cyclically permuting the signs of
\begin{multline}
\left| \arcsin \braket{A_0B_0} + \arcsin \braket{A_0B_1} \right. \\
+ \left. \arcsin \braket{A_1B_0} - \arcsin \braket{A_1B_1} \right| \le \pi.
\label{eq: TLM inequalities}
\end{multline}
This characterisation allows to randomly generate a set of correctly labelled behaviours that can be used to train an ML model. Unfortunately, a similar description of the boundary for the entire quantum set is lacking, which prevents a straightforward classification of behaviours as quantum or not quantum. Nonetheless, before addressing the topic of data classification, which is discussed in detail in Sec.~\ref{sec: data generation}, let us briefly introduce the fundamental concepts and notation of the field of ML.

\subsection{Machine learning in a nutshell}

Machine Learning is an interdisciplinary field at the intersection of computer science and statistics that is dedicated to developing methods that allow computers to learn and make decisions based on data \cite{Bishop06}. A fundamental aspect of ML is its ability to build predictive models that can analyse data and identify patterns or trends. Such ability is advantageous in applications where traditional analysis methods fail to provide reliable predictions.

Conventionally, ML models are split into three categories, namely supervised learning, unsupervised learning, and reinforcement learning. For the purposes of this work, we will concentrate on supervised learning, i.e., a technique which is characterised by the use of labelled datasets to train the models \cite{Hastie09}. These datasets consist of input-output pairs, where the model is provided with input data, often referred to as \textit{features}, along with the correct \textit{labels}, allowing it to learn how to predict the output from the input.

There are two main types of tasks in supervised learning, which basically differ in the nature of the output: regression tasks deal with continuous output, e.g., predicting house prices, stock market trends, or temperature forecasts, while classification tasks deal with predicting discrete outputs, typically categories or classes. Our work presents the quantum correlation problem as a classification task in which the features are given by the CHSH behaviour, as in Eq.~\eqref{eq: correlators representation}, and the labels consist of a binary class dividing each feature as being quantum or not quantum.

From an operational point of view, constructing an ML model typically involves a few elementary steps, namely data preprocessing, model selection, training and testing. In our particular context, this paradigm shifts slightly as we do not use pre-existing data but rather generate it ourselves, a process detailed in the subsequent section. Following data preprocessing, further steps involve selecting and training a given ML model. The choice of model depends on the nature of the task (e.g., regression, classification), the characteristics of the data and the expected difficulty of the task. Common models in supervised learning include linear regression, logistic regression, support vector machines, and decision trees, among others. More complex problems may require advanced techniques such as neural networks.

Once the data and model are fixed, the subsequent step is to divide the dataset into training and testing sets. This division is critical for evaluating the model’s performance on unseen data, which is a proxy for how it will perform in real-world scenarios. A common practice is to allocate a larger portion of the data for training (e.g., 70-80\%) and a smaller portion for testing (e.g., 20-30\%). Here, due to the nature of the chosen models, we also include a validation set, splitting the generated data into validation, training and testing sets, following a 15-70-15 proportion, respectively. After the data is split, the training process involves using the validation set to tune the hyperparameters of the model. This process is iteratively performed to find the optimal parameter values that minimise a predefined loss function, which quantifies the difference between the predicted and actual values.\footnote{It should be noted that the term ``training'' referred here is what in statistics is referred to as ``fitting'', i.e., the optimisation of parameters in statistical models (like linear regression). ML's ``training'' not only mirrors this aspect but also includes the feature of boosting model performance through data exposure.} The split between training and validation is crucial as it ensures that the evaluation obtained through the test set consists of a realistic estimate of the model's performance, which brings us to the last step of model development. 

The assessment of the model involves using evaluation metrics applied to the test set, which may vary based on the type of task. For instance, accuracy, precision, recall, and the F1 score are commonly used for classification tasks, while mean squared error or mean absolute error are used for regression tasks. The evaluation phase not only provides insights into the model’s performance but also helps in diagnosing issues such as overfitting, where the model performs well on the training data but poorly on previously unseen examples.

Although the description of the model development process may appear linear based on the sequence of outlined steps, in reality it rarely follows this path exactly. This divergence is evident in our approach, where some steps are skipped or modified. For this reason, we emphasise that the last paragraphs are not intended to describe our model construction specifically, but to provide a basic understanding of ML to a non-specialised audience. In the following, we introduce the two classes of models employed in this work, namely support vector machines and dense neural networks, as well as discuss why we consider this choice suitable for the problem of quantum correlations.

\subsubsection{Support vector machines}

Support vector machines are a class of supervised learning models used primarily for classification, but also for regression tasks \cite{Noble06}. Initially, they were designed to solve linearly separable problems \cite{Vapnik06}, where the data can be split by a flat boundary, that is, a straight line, a plane or a hyperplane, depending on the dimension of the features. In this sense, the goal of the SVM algorithm is to find the optimal hyperplane separating classes, which is defined as the one maximally distant from data points of either class. It follows then that the model's performance strictly depends on the data points closest to this optimal hyperplane, denoted as \emph{support vectors}, which preserves the method's efficiency as it scales to higher dimensions.

The real breakthrough in the utility of SVMs came with the realisation that they could also be adapted to solve non-linear problems through the kernel trick \cite{Boser92, Cortes96}. This approach involves applying a kernel function to map the original non-linearly separable data into a higher-dimensional space, where it becomes linearly separable. Common kernel functions include the polynomial, radial basis function (RBF), and sigmoid kernels. By using this trick, SVMs can efficiently classify data that is non-linearly separable in its original feature space, which makes it a natural candidate for the problem tackled in this work. In other words, SVMs not only exhibit good scalability, which is desirable for Bell scenarios where the dimensionality of correlations is typically high, but they also handle the non-linear aspects of the boundary of the quantum set. This contrasts with linear ML models such as Bayesian or logistic regression, which generally fail in such contexts.

\subsubsection{Neural networks}

Inspired by the neural architecture of the human brain, neural networks \cite{Haykin00} are a sophisticated class of ML models consisting of interconnected nodes, also known as \textit{neurons}, which collectively process and interpret data inputs. The advantage of such models lies in their versatility to learn and adapt to a wide variety of tasks, ranging from image recognition to natural language processing.

In this work, we focus on one of the simplest types of neural networks denominated as multi-layer perceptrons (MLPs) \cite{Goodfellow16}, a type of feedforward neural network. The ``feedforward'' aspect refers to the unidirectional flow of data, i.e., the data supplied to a given neuron does not return to previous neurons, which guarantees a non-cyclical flow of information with no loops or backward connections. This architecture is particularly beneficial for tasks where the current output depends solely on the current input, with no dependence on the context provided by previous inputs.

\begin{figure}[t!]
    \includegraphics{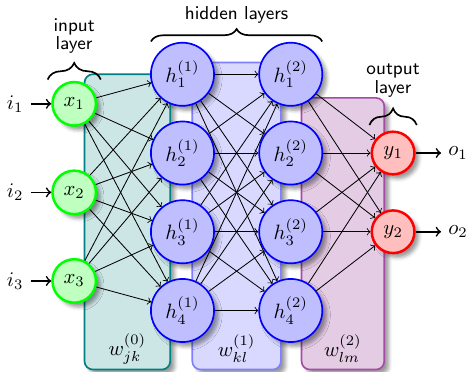}
    \caption{A typical MLP diagram showing two hidden layers. The initial layer consists of three neurons $x_j$ in which the inputs $i_j$ are passed, for $j=1,2,3$. Each input neuron is connected to the $k$-th neuron $\smash{h^{(1)}_k}$ in the first hidden layer through the set of weights $\smash{w^{(0)}_{jk}}$, for $k=1,2,3,4$. Similarly, the neurons in the first hidden layer are connected to $\smash{h^{(2)}_l}$ through the weights $\smash{w^{(1)}_{kl}}$, for $l=1,2,3,4$. The second hidden layer ends up in the output layer via $\smash{w^{(2)}_{lm}}$, where $m = 1,2$. At the end of the process, $y_1$ and $y_2$ yield output $o_1$ and $o_2$, respectively.}
    \label{fig: neural network example}
\end{figure}

In this sense, MLPs exemplify this feedforward structure, featuring layers of interconnected neurons divided into an input layer, one or more hidden layers and an output layer, as illustrated in FIG.~\ref{fig: neural network example}. Each layer in an MLP is structured in a way which is commonly referred to as \emph{dense}, where every neuron in a given layer is connected to all neurons in the next layer through a set of adjustable weights. During its operation, each neuron first computes a weighted sum of its inputs and then applies an activation function. The activation function is crucial as it introduces non-linearity in the model, which allows the network to capture complex patterns from the data. Common activation functions include the sigmoid, hyperbolic tangent, and ReLU (Rectified Linear Unit) \cite{Glorot11, Ramachandran17}.

The learning ability of MLPs relies on their method of iteratively refining weights during training. At this stage, the network undergoes a process denoted as forward pass, where each training example is fed through the network from the input to the output layer, generating a prediction. Then, the loss calculation phase assesses the accuracy of these predictions comparing them to the true values with a loss function, which quantifies the error in the previous step. The network then computes the gradient of the loss function with respect to each weight in a process known as backward pass or backpropagation. In the last step, an optimisation algorithm, often gradient descent, adjusts the weights to minimise the loss function. Completing all of these steps for all training examples constitutes one epoch, marking a full cycle through the training data. After multiple epochs, the resulting model should be able to yield refined predictions through the output layer, where neurons are configured to match the specific outputs needed. For classification tasks, the number of neurons in this layer typically corresponds to the number of classes the model should predict, whereas, for regression, a single neuron produces a continuous value. In other words, the goal of the output layer is to transform the outputs of the last hidden layer into a format suitable for the problem.

In summary, MLPs are versatile tools, being well-suited for a wide range of problems, including the one addressed in this work. Additionally, Ref.~\cite{Amos17} points out that if all activation functions are convex and non-decreasing, and all weights are non-negative, the output is guaranteed to be a convex function of the inputs. Such an observation can be applied to MLPs by simply constraining the weights to be non-negative, since most activation functions are already convex and non-decreasing. Here, we use this idea to exploit the convexity of our problem and ensure that the MLP model produces a convex output. Nonetheless, before we enter into the specifics of fitting the model, as discussed in Sec.~\ref{sec: fitting the model}, let us shift our focus to the process of data generation.

\section{Data generation} \label{sec: data generation}

\noindent Model and data are the two main components of a typical ML problem. Distinct models have different computational capabilities and requirements, but most are generally adaptable for various purposes. Naturally, some models will perform better than others, in terms of accuracy and efficiency, given a particular dataset. For this reason, part of the challenge in the search for an ML solution consists of deciding which tool is better suited for the data at hand, which is usually a constraint determined by the problem. However, in the scenario considered here, there is no data available in advance, leaving the only alternative to generate it by ourselves. As this is not a trivial task, especially in distinguishing behaviours between quantum and not quantum, the following subsections are devoted to this discussion, presenting a few different ways to sample and classify non-signalling behaviours.

But before we go any further, let us point out that producing training data may seem counterintuitive here, as we need to classify behaviours first and this is also the task we want our model to perform. Nonetheless, it is still worth to look for an ML model for two reasons. Firstly, an ML model would effectively combine the strengths of the methods we use to generate data, such as see-saw and NPA, consolidating them into a single, unified approach. Secondly, once trained, the resulting model has the potential to be more efficient in terms of computation time and memory than the original methods. This makes exploring the development of a well-trained ML model from a relatively small set of classified examples highly worthwhile.

\subsection{Uniform sampling} \label{subsec: uniform sampling}

In our first approach to generate data, we create a uniform sample within the $\NC\SC$ polytope. This is achieved using an algorithm called \emph{hit-and-run}, which employs a random walk starting from an interior point, as implemented in Ref.~\cite{GitHub18}. The hit-and-run method enables the generation of uniformly distributed points across any convex bounded shape where the boundaries are known analytically. The sample is produced by inputting the hyperplane representation of $\NC\SC$ and a starting point, which we chose to be its geometric centre, corresponding to the origin of the coordinate system in both 4 and 8-dimensional variants of the CHSH scenario. For the correlation space, the hyperplane representation of $\NC\SC$ is given by the facets of a cube in dimension 4, whereas the same representation for the entire scenario is expressed by the non-signalling facets, i.e,
\begin{multline}
(-1)^{a + b + 1}\braket{A_xB_y} + (-1)^{a} \braket{A_x} + (-1)^{b} \braket{B_y} \le 1, \\
\forall~a,b,x,y,
\label{eq: positivity facets}
\end{multline}
where, here, the outcomes are labelled as $a, b\in \{0,1\}$.

\begin{figure}[t]
    \centering
    \includegraphics[width=8.6cm]{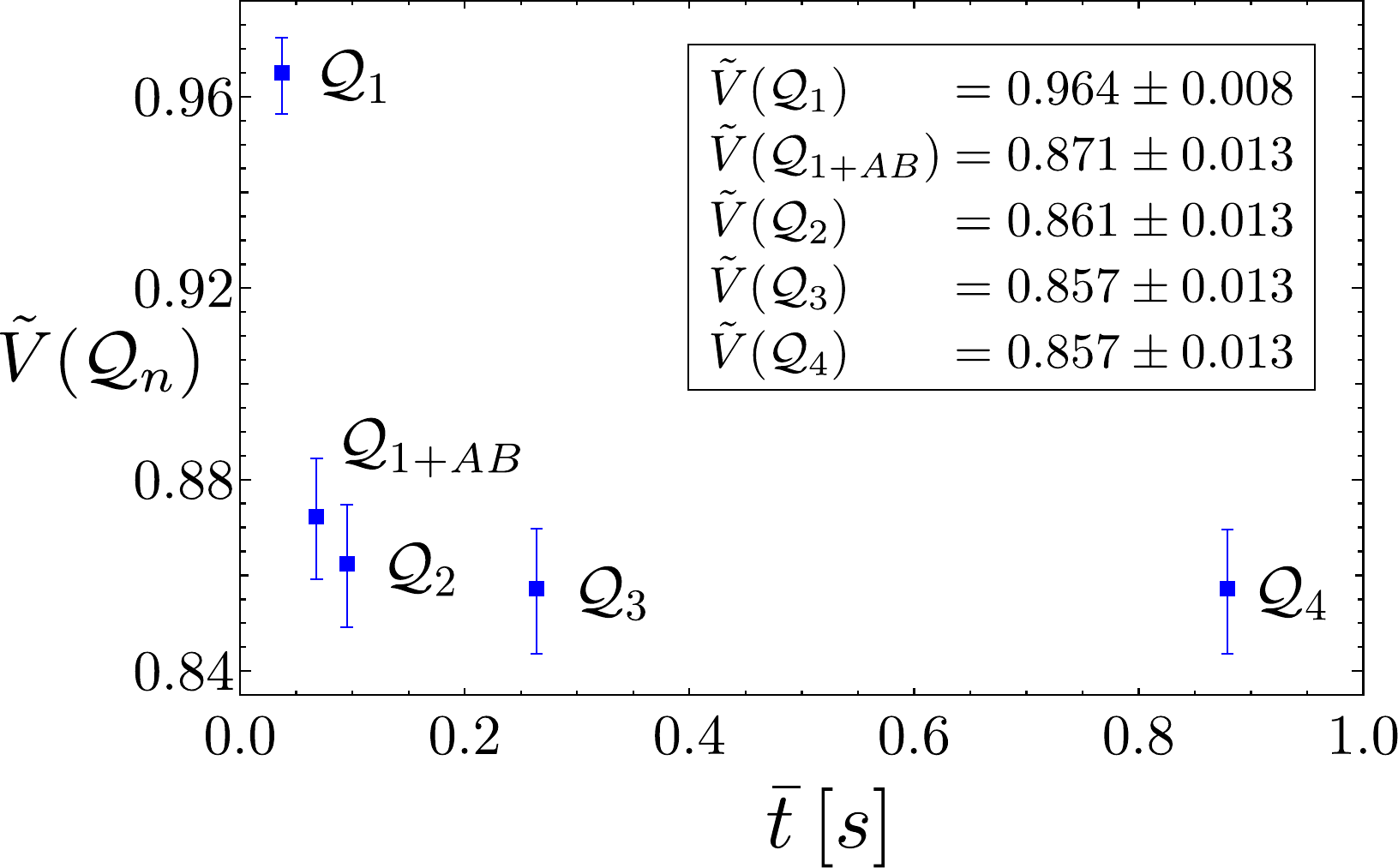}
    \caption{Relative volumes of $\QC_n$ vs. average computation time $\bar{t}$. This plot was produced by computing the volume ratios of $\QC_n$ with respect to $\NC\SC$ using a simple Monte Carlo method. Briefly, a sample of points uniformly distributed in $\NC\SC$ is generated, and the membership is solved for each point. Note that here, the volumes of $\QC_n$ are considered regarding the 8-simplex violating one of the CHSH facets, i.e., the sample produced is composed only of non-local points. For each NPA level, an initial sample of 10\,000 points is created and classified, a procedure repeated 50 times. The average CPU time $\bar{t}$ per point is shown in the $x$-axis. It is possible to observe that levels higher than $1+AB$ provide a small improvement towards the actual volume of $\QC$ at the cost of a significant increase in CPU time. The CPU clock speed used is 2.9 GHz (base) and 3.9 GHz (boost).}
    \label{fig: relative volumes}
\end{figure}

Once the sample is generated, the classification of points as quantum or not quantum is straightforward for correlations only, as the first level of NPA is enough to solve the membership problem. Nonetheless, for the entire CHSH scenario there is no analogous method allowing to easily classify correlations points, and the best alternative is to approximate $\QC$ with the superset $\QC_n$ defined by the $n$-th level of NPA. The drawback is that the SDP associated with the membership of a given level becomes costly as $n$ increases, creating a compromise between the quantity of resources to be used and the quality of the approach. In FIG.~\ref{fig: relative volumes}, this question is addressed with the comparison between the relative volumes of the first NPA supersets, showing an advantage in terms of computation time to level $1+AB$. Although NPA converges when $n \rightarrow \infty$, the quality of the approximation for large $n$ does not improve drastically when compared to previous levels. Furthermore, the execution time increases in a way that makes the computation infeasible for datasets larger than $10^5$ points, even for level $n=5$. In this way, we aim to create datasets for training ML models to decide whether a given behaviour belongs to $\QC_{1+AB}$ or not, which, in good approximation, would tell us if it belongs to $\QC$.

Also, it is worth noting that the volumes spanned by $\QC$ and $\NC\SC$ in the CHSH scenario are almost the same, making the volume ratio close to one\footnote{For the correlation space, this number is known, and it is approximately $0.925$ \cite{Cabello05, Wolfe12}.}. A consequence of this feature is that a model that trivially predicts every point to be local can be considered good if accuracy with respect to a uniformly distributed sample is taken as a figure of merit to evaluate performance. Moreover, it is likely that an unbiased loss function would prioritise the larger class in the problem. Here we consider three ways in which this data imbalance can be handled. First, we use balanced loss functions, like the balanced binary cross entropy, for binary classification, and balanced figures of merit for further evaluation of the models. Secondly, we generate a balanced sample for the two classes, which would make a model that trivially predicts all points as belonging to one class perform badly in terms of accuracy. Lastly, we introduce a technique that samples the training data only near the boundary between classes. Intuitively, this seems a smart way to proceed, since focusing the training data near the boundary naturally leads to a more difficult problem, as points far away from the boundary are easier to predict. Consequently, achieving greater accuracy in such a model would imply a decision boundary that resembles the true boundary more accurately. To describe this idea in more detail, we move on to the next subsection.

\subsection{Sampling near the quantum boundary} \label{subsec: sampling near the boundary}

Briefly speaking, the idea behind this method is to find a correlation point on the boundary of $\QC$, and, from this point, generate a pair of training points belonging to each of the classes. For the correlation space, since $\QC_1 = \QC$, the procedure is straightforward and is implemented with the first level of NPA. To do so, we begin with the origin in $\mathbb{R}^4$ as the starting point, and draw a random unit direction $\bf u$ in the 4-sphere. Then, by solving the SDP,
\begin{align}
\begin{split}
\max \quad & \lambda \\
\text{s.t.} \quad & \lambda\bf u \in\QC_1,
\end{split}
\label{eq: NPA SDP}
\end{align}
we can find a solution $\lambda^*$ which produces a point $\mathbf{p}_b = \lambda^* \bf u$ in the boundary of $\QC_1$. The training points are then created from $\mathbf{p}_b$ by considering the pair
\begin{align}
\mathbf{p}_\pm = (1 \pm \epsilon)\, \mathbf{p}_b, 
\label{eq: training pair}
\end{align}
where $\epsilon$ is a fixed offset. This ensures that the training set is balanced by construction, since the classification is done according to the offset sign, i.e., $\bf p_-$ is always quantum, while $\bf p_+$ is not. Also, through direct numerical verification, it is possible to check that the classification in Eq.~\eqref{eq: training pair} compared to the analytical TLM conditions results in correctly classified samples for offsets greater than $10^{-10}$, which is sufficient for our numerical purposes.

As for the entire quantum set, the procedure is analogous, beginning by choosing a direction in $\mathbb{R}^8$. Since there is no known finite level of NPA that corresponds to the actual quantum set, $\QC_{1+AB}$ replaces $\QC_1$ in Eq.~\eqref{eq: NPA SDP}. With the method detailed in the following subsection, we show in Appendix~\ref{app: Q1AB evaluation} that this approximation generates training points that are mostly correctly classified when $\epsilon \ge 10^{-3}$.

It is also worth mentioning that, while proximity of the training data to the NPA superset is ensured for small $\epsilon$ in Eq.~\eqref{eq: training pair}, the actual distance of the training points may vary depending on $\bf u$. Additionally, since $\bf u$ will be generally tilted with respect to the direction normal to the boundary in the vicinity of ${\bf p}_b$, the density of training points will vary from one region to another. In an attempt to mitigate these issues, we initially devised two variations to the method described in Eqs.~\eqref{eq: NPA SDP} and \eqref{eq: training pair} for the correlation space models. The first of them involved performing a second SDP to compute the distance between $\mathbf{p}{\pm}$ and $\QC_1$, thereby ensuring a uniform distance for all training points to the set. The second variation involved pre-selecting the directions $\mathbf{u}$ to guarantee a uniform distribution of points $\mathbf{p}_b$ on the hyper-surface of $\QC_1$. Nonetheless, although the training data produced by these variations differed slightly, their effects on the final ML models were negligible and boiled down to fluctuations in the values of the figures of merit. As a result, we decided to retain only the simplest model, which is outlined in Eqs.~\eqref{eq: NPA SDP} and \eqref{eq: training pair}.

\subsection{Steered see-saw} \label{subsec: steered see-saw}

Similarly to the see-saw optimisation, the following method also consists of an iterative SDP technique. But here, instead of maximising a given linear functional, we compute the distance of a point $\mathbf{p}_b$, on the boundary of one of the NPA supersets, to the set of points admitting a quantum realisation of local dimension $d$, denoted as $\QC_{d \times d}$. That is, the central idea essentially mirrors the task of computing the distance between a point and a convex set. However, while $\QC_{d \times d}$ is not always convex \cite{Pal09}, the procedure is divided so that each step is cast as a convex problem. 

In the first of these steps, we randomly select observables $\tilde{A}_x$ and $\tilde{B}_y$, for $x,\, y \in \{0,1\}$, acting in a Hilbert space of local dimension $d$, to write
\begin{align}
\begin{split}
\min_{\rho} \quad & || \mathbf{q} - \mathbf{p}_b || \\
\text{s.t.} \quad & \mathbf{q} \in \QC_{d \times d}.
\end{split}
\label{eq: optimise rho}
\end{align}
Here, the symbol $|| \cdot ||$ denotes the 2-norm, and $\bf q$ represents a correlation computed out of the averages over $\tilde{A}_x$, $\tilde{B}_y$ and the variable state $\rho$. By using the solution $\tilde{\rho}$ obtained from Eq.~\eqref{eq: optimise rho}, we then perform an analogous optimisation for the observables of each part alternately. After completing all steps, the resulting distance is compared to a previously defined bound, set at $10^{-7}$. If the distance is smaller than this threshold, the process concludes; otherwise, it iterates for a given number of seeds. That is, this procedure essentially acts as a classification test: $\mathbf{p}_b$ belongs to $\QC$ when the distance falls below the threshold, but no conclusion is drawn if it does not.

\begin{table}[b!]
    \centering
    \setlength{\tabcolsep}{6pt}
    \caption{CPU time for each sampling method in the 8-dimensional CHSH space. The first column specifies the method utilised, while the second column shows the total CPU time to classify an initial sample of $10^4$ points. The first and second rows show the case of a uniformly distributed sample when the classes are unbalanced and balanced, respectively. The sampling strategy is similar for both methods, with the difference that the balanced sample accumulates non-local points until the size of each class reaches half of the desired sample size. The two last rows show the time taken for the methods discussed in Secs.~\ref{subsec: sampling near the boundary} and \hyperref[subsec: steered see-saw]{C}, respectively. It is worth noting that the steered see-saw consists of a classification test in $\QC$, while the other methods use $\QC_{1+AB}$ as reference. For more details regarding the classification with steered see-saw, see Appendix~\ref{app: Q1AB evaluation}.}
    \medskip
    \begin{tabular}{ c | c }
        method & total CPU time \\ [0.5 ex] 
        \hline\hline
        uniform sampling (unbalanced) & 6 min 52 s \\ [0.15 ex]
        \hline
        uniform sampling (balanced) & 43 min 27 s \\ [0.15 ex]
        \hline
        sampling near the boundary & 3 min 23 s \\ [0.15 ex]
        \hline
        steered see-saw & $\gtrsim$ 158 days \\ [0.15 ex]
        \hline
    \end{tabular}
    \label{table: time per method}
\end{table}

When compared to the standard see-saw, the approach outlined here has a couple of advantages. The first is that the optimisation is not carried out in relation to a particular Bell functional, but to a direction in the coordinate system, allowing to explore non-exposed points on the boundary of $\QC$. Second, even though the membership test may be inconclusive, for a sufficiently large dimension and number of seeds, it can be interpreted as negative, since $\QC_{d^\prime \times d^\prime} = \QC$, for $d^\prime \ge 16$ \cite{Donohue15}. On the other hand, the feasibility of this procedure for applications involving a sufficiently large dataset is far from practical. TABLE~\ref{table: time per method} presents a CPU time comparison for the methods discussed in this section to produce a sample of the same size. Clearly, the steered see-saw has a significantly larger cost than the previous methods. While the other procedures may typically require a couple of SDPs to solve the membership problem, the latter method subjects a single point to a series of SDPs, in addition to the need to iterate over $d$. The result is that, though we can use the steered see-saw in the small correlation space, producing a training set with this method for the entire 8-dimensional space of the CHSH scenario is computationally challenging.

Nonetheless, despite the fact that we cannot directly use this approach for our ML task, it can still be applied for some other questions. In Appendix~\ref{app: Q1AB evaluation}, we assess the quality of the approximation made in Sec.~\ref{subsec: sampling near the boundary} when using $\QC_{1+AB}$ as the reference set for classification. In short, we draw a sample of $10^4$ unit directions in $\mathbb{R}^8$ and use Eq.~\eqref{eq: NPA SDP} to determine an optimal $\lambda$ for $\QC_{1+AB}$. The results are then compared with the steered see-saw using $d = 6$ and 50 seeds. The outcome is that for around 99.46\% of the directions starting from the origin in $\mathbb{R}^8$, the points produced via Eq.~\eqref{eq: training pair} are correctly classified when $\epsilon \ge 10^{-3}$, showing that resolving $\QC$ from $\QC_{1+AB}$ numerically is difficult. Given that ML models are intrinsically probabilistic, this implies that even if a training sample using $\QC$ as a reference is produced, the results obtained with such a dataset would be no different from those produced by using $\QC_{1+AB}$ as a reference.

\section{Fitting the model} \label{sec: fitting the model}

\noindent In this section, we discuss the training details of the chosen ML models, as well as present a performance analysis based on the evaluation metrics available for each. In the first part, we show the results of the fitting in the correlation space, subsequently moving to the entire scenario in the second part. Each subsection is divided according to one of the two models considered, namely SVMs and NNs.

It is worth to note that aside from the aforementioned models, we have also experimented with other types of models which can be found in Python's Scikit-learn library \cite{Pedregosa11}, such as Gaussian Processes, Naive Bayes, and Decision Trees. However, we have decided to discontinue some of the original approaches due to distinct reasons. On the one hand, Decision Trees were found to lack stability, with slight variations in training data leading to drastic model changes, making it difficult to be reproduced. Moreover, they were sensitive to the choice of coordinate system, i.e., given that the correlations allow for arbitrary choices of coordinates, we cannot predict in advance which coordinate system will lead to the most effective tree structure. In contrast, Naive Bayes and Gaussian Processes demonstrated to be more reliable but were discarded due to Naive Bayes' inferior accuracy and the extensive training time required by Gaussian Processes. Unlike SVMs, which are powered by the C++ library LIBSVM \cite{Chang11} in Scikit-learn, Gaussian Processes are fully implemented in Python. While this implementation offers great flexibility and ease of use, it may not achieve the same performance as systems optimised in lower-level languages, especially for large-scale problems. Therefore, it clearly became apparent that SVMs and NNs are the most promising directions, which we cover in the following.

\subsection{Benchmark in the correlation space}

Our initial interest in this particular subset comes from the fact that here the boundary of $\QC$ is known to be given by the TLM inequalities, also coinciding with the boundary of the first NPA superset $\QC_1$. In what follows, we explore the performance of SVMs in producing a decision boundary that approximates that of $\QC$, i.e., in solving the problem of classifying non-signalling points as quantum vs. not quantum.

\subsubsection{SVMs on correlations only}

As mentioned above, in the development of the models discussed here, the Scikit-learn library was used as the main modelling tool, employing the RBF kernel among the available kernel options. Based on the datasets appearing in the three first rows of TABLE~\ref{table: time per method}, we developed three distinct SVM models, which we refer to as unbalanced, balanced, and offset models, respectively. In each case, a sample of $10^4$ points was created following the standard split of 70\% for the training set and 15\% for the test and validation sets each. In all three models, the adjustment of the SVM hyperparameters was performed through a grid search in the hyperparameter space, selecting those values that maximise accuracy in each validation set. For the offset model, in addition to the hyperparameter search, it was also necessary to define the offset parameter $\epsilon$ used to produce the sample. In this case, we started tentatively with $\epsilon=10^{-2}$, obtaining the accuracies shown in TABLE~\ref{table: accuracies wider gap}.

\begin{table}[b!]
    \centering
    \setlength{\tabcolsep}{5.5pt}
    \caption{Accuracies for the correlation SVM models. The second and third columns contain the accuracies computed using the testing and training sets as a reference, respectively. The last column shows the accuracies when the models are tested using the unbalanced test set, i.e., when the test set is taken as a sample of points uniformly distributed within the correlation space of $\NC\SC$.}
    \small
    \begin{tabular}{c|c|c|c}
        model & test acc. & train acc. & unbalanced acc.\\ [0.5 ex]
        \hline\hline
        unbalanced & 0.9926 & 0.9934 & --- \\ [0.15 ex]
        \hline
        balanced & 0.9840 & 0.9901 & 0.9796 \\ [0.15 ex]
        \hline
        offset ($\epsilon=10^{-2}$) & 0.9307 & 0.9961 & 0.9800 \\ [0.15 ex]
        \hline
    \end{tabular}
    \label{table: accuracies wider gap}
\end{table}

Initially, we considered the accuracy measures relative to three reference sets: the training and testing sets of each model, to assess generalisation within the same data type, and the unbalanced test set, to determine generalisation across the entire $\NC\SC$ polytope. Despite the good performance of the three models, the small variations observed in TABLE~\ref{table: accuracies wider gap} make it difficult to evaluate the influence of each dataset and, specifically, the effect of $\epsilon$ on the quality of the predictions nearby the boundary. To address this issue, we included an additional accuracy measure, denoted here as \emph{spread test}, which also involves sampling the reference set near the boundary of $\QC$.

\begin{figure*}[t!]
    \centering
    \includegraphics[width=\textwidth]{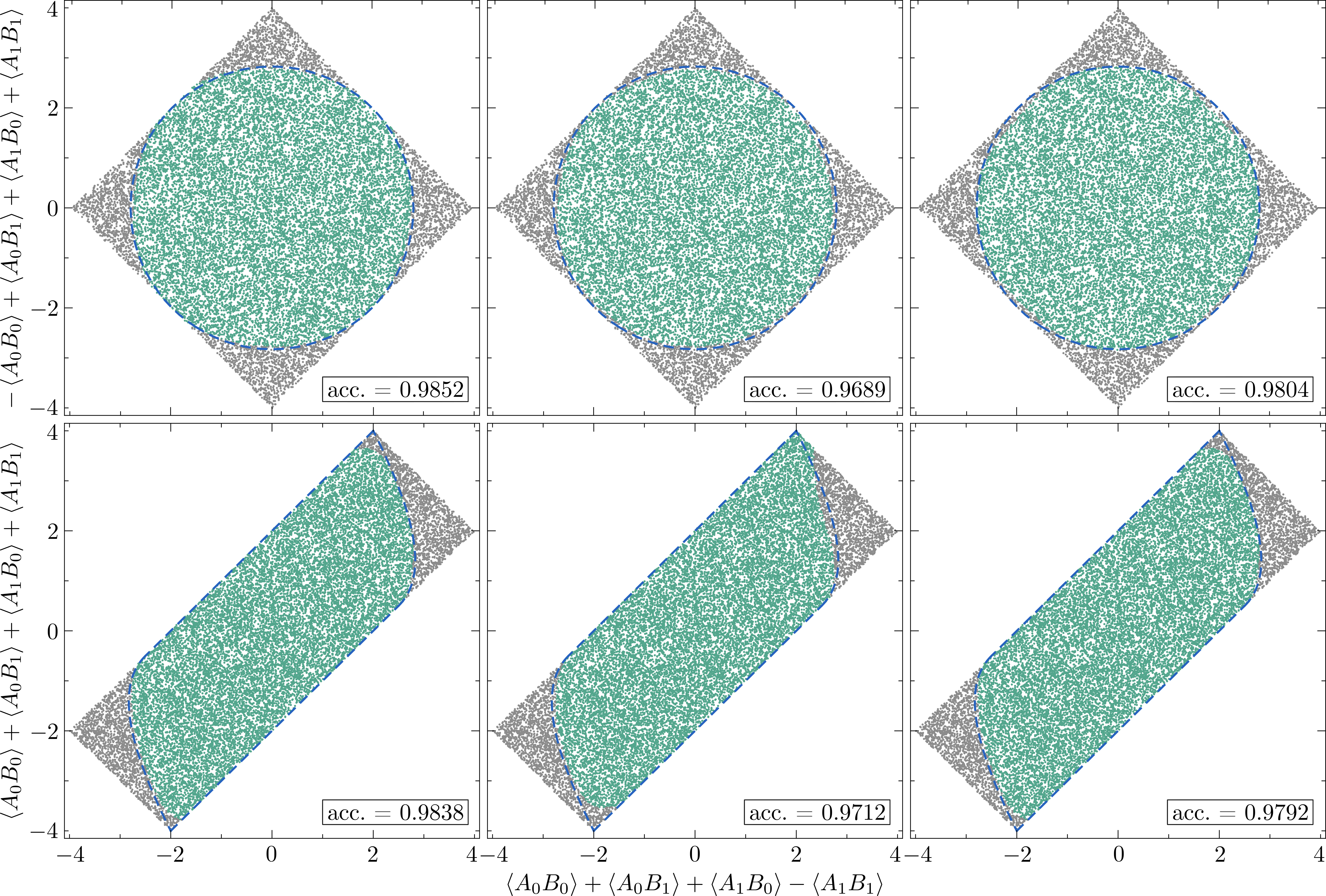}
    \caption{Correlation slices for the unbalanced (right), balanced (centre) and offset (left) SVM models with $\epsilon = 10^{-3}$. The upper and lower slices were introduced in Refs.~\cite{Branciard11} and \cite{Goh18}, respectively. The green dots depict the quantum class, while gray represent the points classified as outside $\QC$. The dashed line depicts the true boundary given by the TLM condition. In each plot, $2 \times 10^4$ points were generated and classified; the accuracies are indicated in the lower corner for each slice.}
    \label{fig: 4D SVM slices}
\end{figure*}

To generate this reference set, we adopted a different sampling method from that discussed in Sec.~\ref{subsec: sampling near the boundary}. To put it briefly, the distinction lies in how the initial boundary points are collected; rather than being sampled in a random unit direction, they are uniformly distributed along the hyper-surface defined by the boundary of $\QC_1$. This is achieved by creating a uniformly distributed sample within a shell of very small thickness that encloses $\QC_1$. The directions of each point in this sample are then used by the method in Sec.~\ref{subsec: sampling near the boundary} to find approximations on the boundary of $\QC_1$. Lastly, each boundary point is shifted as $[1 + \NC(0, \sigma)]\, \mathbf{p}_b$, where $\NC(0, \sigma)$ is a normal distribution centred at zero and with standard deviation $\sigma$. Labels are assigned by directly checking the sign of $\NC(0, \sigma)$.

\begin{table}[b!]
    \centering
    \setlength{\tabcolsep}{7pt}
    \caption{Accuracies produced for the spread test. Columns two and three present the values obtained when the test corresponds to boundary points shifted by a normal distribution $\NC(0, \sigma)$ with $\sigma=10^{-2}$ and $10^{-3}$, respectively. The smaller the spread of the distribution, the closer the test points are to the boundary of the set.}
    \small
    \begin{tabular}{ c | c | c }
        model & spread acc. ($\sigma=10^{-2}$) &  ($\sigma=10^{-3}$) \\ [0.5 ex] 
        \hline\hline
        unbalanced & 0.5430 & 0.5030 \\ [0.15 ex]
        \hline
        balanced & 0.5426 & 0.5027 \\ [0.15 ex]
        \hline
        offset ($\epsilon=10^{-2}$) & 0.8318 & 0.5920 \\ [0.15 ex]
        \hline
    \end{tabular}
    \label{table: new accuracies}
\end{table}

The spread test results are detailed in TABLE~\ref{table: new accuracies}. The offset model exhibits good performance at $\sigma = 10^{-2}$, decreasing at $\sigma = 10^{-3}$, whereas the balanced and unbalanced models consistently perform poorly, given that 0.5 can be achieved with a model that guesses the labels randomly. Additionally, this test serves to highlight how the $\epsilon$ value influences prediction accuracy near the boundary. More specifically, with an offset at $\epsilon=10^{-3}$, the accuracies listed in TABLE~\ref{table: test train hard accuracies} were observed. Although this smaller $\epsilon$ leads to a mild overfitting, as shown by the gap between training and testing accuracies, it improves the resolution at the boundary without disturbing the model's ability to predict other regions. FIG.~\ref{fig: 4D SVM slices} illustrates the agreement across the unbalanced, balanced, and offset (with $\epsilon=10^{-3}$) models in two correlation slices, where the decision boundary aligns closely with the analytical boundary shown by the dashed line.

\begin{table}[t!]
    \centering
    \setlength{\tabcolsep}{8pt}
    \caption{Accuracies for the offset SVM model with $\epsilon=10^{-3}$. Column one contains a summary of all accuracies appearing in TABLES~\ref{table: accuracies wider gap} and \ref{table: new accuracies}, and column two displays their values. Compared to $\epsilon=10^{-2}$, although performance was reduced for the training and test sets, the values for the unbalanced sample improved, as well as for the spread test.}
    \small
    \begin{tabular}{ c | c }
        acc. measure & value \\ [0.5 ex]
        \hline\hline
        test acc. & 0.7353 \\ [0.15 ex]
        \hline
        train acc. & 0.8128 \\ [0.15 ex]
        \hline
        unbalanced acc. & 0.9912 \\ [0.15 ex]
        \hline
        spread acc. ($\sigma=10^{-2}$) & 0.8923 \\ [0.15 ex]
        \hline
        spread acc. ($\sigma=10^{-3}$) & 0.6586 \\ [0.15 ex]
        \hline
    \end{tabular}
    \label{table: test train hard accuracies}
\end{table}

To conclude, it remains to be mentioned that although the smaller offset leads to a better resolution at the boundary, it cannot be further improved. That is, in our experiments, as $\epsilon$ diminishes further, finding a good fitting becomes increasingly hard. The gap between training and testing accuracies increases while the performance on uniformly distributed data declines, suggesting that there is a limit to how much the offset value can be reduced while maintaining the ability to accurately classify correlation points.

\subsubsection{NNs on correlations only}

Apart from the SVM models mentioned above, we also approached the problem with the implementation of a feedforward neural network. This model was also executed in Python, using TensorFlow \cite{Abadi16} and Keras \cite{Chollet15}. We considered here the same datasets as in the previous discussion for SVMs. For each of them we configured a network with $64$ neurons in the input layer, $16$, and $4$, respectively, for the first and second hidden layers, and two neurons in the output layer, corresponding to each class in our problem. The chosen activation function is ReLu for all neurons except for the last one, for which we chose Softmax. Note that, since we are dealing with two classes, we could instead use a single neuron in the last layer activated with a Sigmoid function, which is mathematically equivalent in terms of activation. We observed, however, that the first option results in more accurate models. We complemented this basic architecture with an extra feature: Since the boundary between classes is known to be convex, we enforced this constraint on the decision boundary by forcing the neuron weights to be non-negative, as suggested in Ref.~\cite{Amos17}. Nonetheless, we observed the impact of this constraint to be negligible in terms of model performance or training time. The architecture we chose for our network is the result of a trial and error approach, in which we looked for the simplest network that can provide a good model, as measured by our figures of merit.

Regarding the training of the network, a common choice of loss function for classification problems is the Binary Cross Entropy (BCE) function. We have found, however, that a variation of it known as Focal Loss results in models with a better performance. The Focal Loss function is defined in \cite{Yi18} as $FL(p) = -\alpha(1-p)^\gamma\log(p)$, where parameter $\alpha$ is introduced to compensate class imbalances, and the \emph{focusing} parameter $\gamma$ diminishes the weight of easy-to-classify behaviours on the loss function, so that the training is carried mostly on hard examples, which lie near the boundary between classes. We observe that a suitable choice of the value of these hyperparameters $(\alpha = 10^{-2}, \gamma = 2)$ results in a noticeable improvement in terms of accuracy for the models trained with data uniformly sampled inside $\NC\SC$.

\begin{figure}[t!]
    \centering
    \includegraphics[width = 8.6cm]{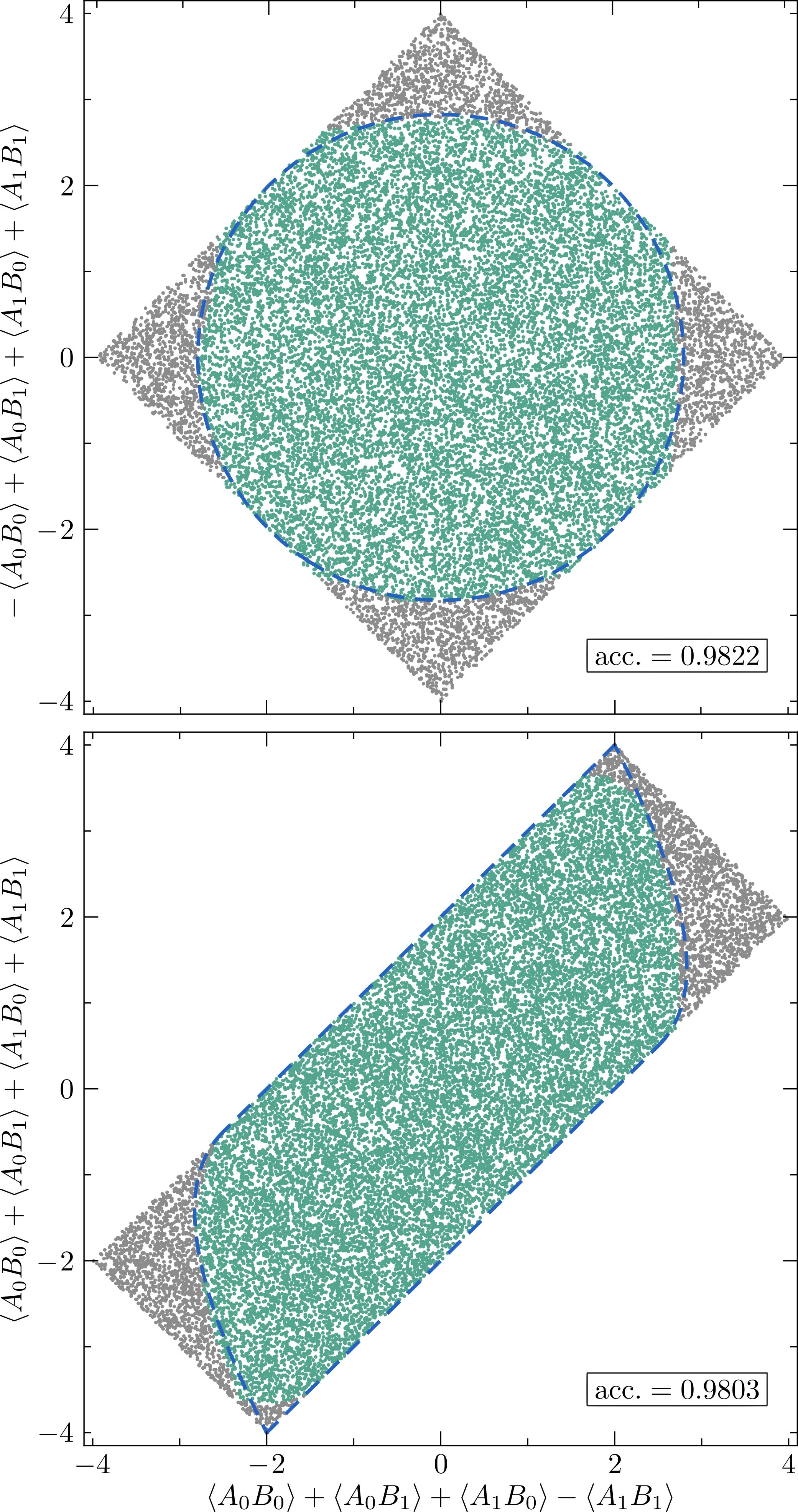}\\
    \caption{Performance of the neural network model trained with the unbalanced dataset on two slices of the correlation space. The slice on top is defined by two orthogonal PR boxes, whereas the one on the bottom is defined by the relation $\braket{A_0B_0} = \braket{A_0B_1} = \braket{A_1B_0}$. The panels depict a set of $2 \times 10^4$ points on each slice, classified by the NN.}
    \label{fig: 4D NN plots}
\end{figure}

In terms of evaluation, on the other hand, we decided to introduce a custom metric which is a balanced version of the Binary Accuracy metric, as a figure of merit to assess the performance of the trained model. The reason for the introduction of this custom metric is that it is expected to be better suited for assessing model accuracy in the presence of unbalanced datasets. We observed indeed that this metric can resolve better the differences between trained models, despite their relatively small magnitude ($\approx 1-10\%$). We trained the neural network using the same training sets that gave rise to the unbalanced, balanced and offset (with $\epsilon=10^{-3}$) models described previously, with each set split into training, validation and test. In order to maximise the performance of the final model, we trained up to $10$ models with different random initial values for the neuron weights, out of which we kept the model with higher accuracy. The results of the neural network training are shown in TABLE~\ref{table: NN accuracies}, in which, as before, we show the accuracy values obtained for the test, train and unbalanced test sets. To further illustrate the model performance, we have considered a pair of slices of the CHSH correlation spaces and computed the model accuracy within these slices, which are shown in FIG.~\ref{fig: 4D NN plots}. To avoid repeating similar plots, in FIG.~\ref{fig: 4D NN plots} we only present the results for unbalanced model; for the remaining models we simply show the plot accuracy in TABLE~\ref{table: slice accuracies} obtained for both slices.

\begin{table}
    \centering
    \setlength{\tabcolsep}{5.5pt}
    \caption{Accuracies obtained for the unbalanced, balanced and offset NN models. For all datasets, we trained the neural network using as function loss the Binary Cross Entropy (BCE) and a balanced version of it.}
    \small
    \begin{tabular}{ c | c | c | c }
        model & test acc. & train acc. & unbalanced acc. \\ [0.5 ex] 
        \hline\hline
        unbalanced &  0.99836 & 0.99581 & 0.99836 \\ [0.15 ex]
        \hline
        balanced & 0.98501 & 0.99505 & 0.98306  \\ [0.15 ex]
        \hline
        offset ($\epsilon = 10^{-3}$) & 0.50252 & 0.50328 & 0.80535  \\ [0.15 ex]
        \hline
    \end{tabular}
    \label{table: NN accuracies}
\end{table}

We can see that the models trained with unbalanced and balanced data attain good scores for the training, validation and test sets, with a small difference in favour of the unbalanced data set. Good scores are also observed when assessing the model in the slices we have chosen. These results, however, are not observed for the offset model, for which the score is barely above one half. Interestingly, as shown in TABLE~\ref{table: NN accuracies}, the performance of these models on the slices is above $0.8$ whereas when evaluated on the unbalanced set we find 0.80535, although it should be noted that the latter accuracies can significantly vary depending on the seed values of the weights. We interpret these differences in the results of model evaluation as a reflection of the difference in the amount of easy to classify points across training sets. Nonetheless, we still observe, contrary to our prior intuition, that the training sets consisting of examples near the boundary result in models that are significantly outperformed by those trained with uniformly distributed examples. In order to explore these differences in more depth, we implemented variations to the sampling method described in Sec.~\ref{subsec: sampling near the boundary}. These variations consisted of randomly distributing the examples created near the boundary, following either a uniform or a normal distribution. No appreciable difference in the obtained results was observed, which suggests that the sampling method is suboptimal for training of neural network models.

\begin{table}[t]
    \centering
    \setlength{\tabcolsep}{7pt}
    \caption{Slice accuracies for all presented SVM and NN models in the correlation space. Slice 1 refers to the accuracy obtained in the upper slice (as in FIGs.~\ref{fig: 4D SVM slices} and \ref{fig: 4D NN plots}) defined by the two orthogonal PR boxes. Slice 2 refers to the lower slice as presented for the first time in Ref.~\cite{Goh18}. For each slice, the same sample of size $2 \times 10^4$ was generated and classified by the respective model, as specified in the first columns.}
    \small
    \begin{tabular}{ c | c | c | c }
        \multicolumn{2}{c|}{model} & slice 1 & slice 2 \\ [0.5 ex] 
        \hline\hline
        \multirow{4}{*}{SVM} & unbalanced & 0.9852 & 0.9838 \\ [0.15 ex]
        \cline{2-4}
        & balanced & 0.9689 & 0.9712 \\ [0.15 ex]
        \cline{2-4}
        & offset ($\epsilon = 10^{-2}$) & 0.9256 & 0.9212 \\ [0.15 ex]
        \cline{2-4}
        & offset ($\epsilon = 10^{-3}$) & 0.9804 & 0.9792 \\ [0.15 ex]
        \hline
        \multirow{3}{*}{NN} & unbalanced & 0.9822 & 0.9803 \\ [0.15 ex]
        \cline{2-4}
        & balanced & 0.9972 & 0.9929 \\ [0.15 ex]
        \cline{2-4}
        & offset ($\epsilon = 10^{-3}$) &  0.8846 & 0.8477 \\ [0.15 ex]
        \hline
    \end{tabular}
    \label{table: slice accuracies}
\end{table}

\subsection{The entire CHSH space}

Here, unlike the previous subsection, the models did not follow a similar thread of development that could be universally applied. Initially, our approach involved expanding the feature dimension across the complete CHSH space, transitioning from the classification previously based on level $1$ of the NPA hierarchy to level $1+AB$. Similar to the correlation models, a dataset comprising $10^4$ points was generated and classified. This dataset was then used to feed the models, which performed poorly both in terms of accuracy metrics and slices. To overcome these issues, different strategies were implemented for each case, starting with the SVM models described below.

\subsubsection{SVMs on the entire CHSH space}

Our approach here leverages the local constraints within the CHSH scenario to tailor our datasets to the non-local region of the $\NC\SC$ polytope. This is done firstly by increasing the feature dimension in our earlier data generation methods to cover the whole CHSH space, and then excluding local points. For balanced and unbalanced models, this adaptation is straightforward, as we simply change the polytope from which the points are drawn using the hit-and-run technique. For this purpose, we employ the 8-simplex defined by the eight local vertices which saturate the inequality 
\begin{align}
    \braket{A_0B_0} + \braket{A_0B_1} + \braket{A_1B_0} - \braket{A_1B_1} \le 2,
    \label{eq: classic CHSH}
\end{align}
and the box $\mathbf{p}_\text{PR} = [0,\, 0,\, 0,\, 0,\, 1,\, 1,\, 1, -1]$ which attains the geometrical bound in Eq.~\eqref{eq: classic CHSH}. Given the one-to-one correspondence between PR boxes and local facets in the CHSH scenario, the $\NC\SC$ polytope can be partitioned between the local set and eight disjoint and identical simplexes, which can be mapped to each other by relabelling settings, outcomes, and exchanging parts. Hence, for any non-signalling correlation, we can classify it as quantum if it meets any local constraint, and otherwise, the model trained on the simplex is applied. In addition, since the local polytope occupies about 94.12\% of the total $\NC\SC$ volume \cite{Wolfe12}, each simplex constitutes approximately 0.74\% of the non-signalling volume, thus decreasing the redundancy in the model trained within this region.  As for the offset model, such adaptation is made by filtering a set of random unit directions in $\mathbb{R}^8$, leaving only those intersecting with the facet in Eq.~\eqref{eq: classic CHSH}. As before, this leads to a reduction in the model's training region, since in terms of the total solid angle comprised by the $8$-sphere, the percentage of selected directions is approximately 0.21\%\footnote{To the best of our knowledge, this value has not been previously reported in the literature. An estimate can easily be obtained by following the procedure outlined in the sentence preceding this note.}.

\begin{figure*}[t!]
    \centering
    \includegraphics[width=\textwidth]{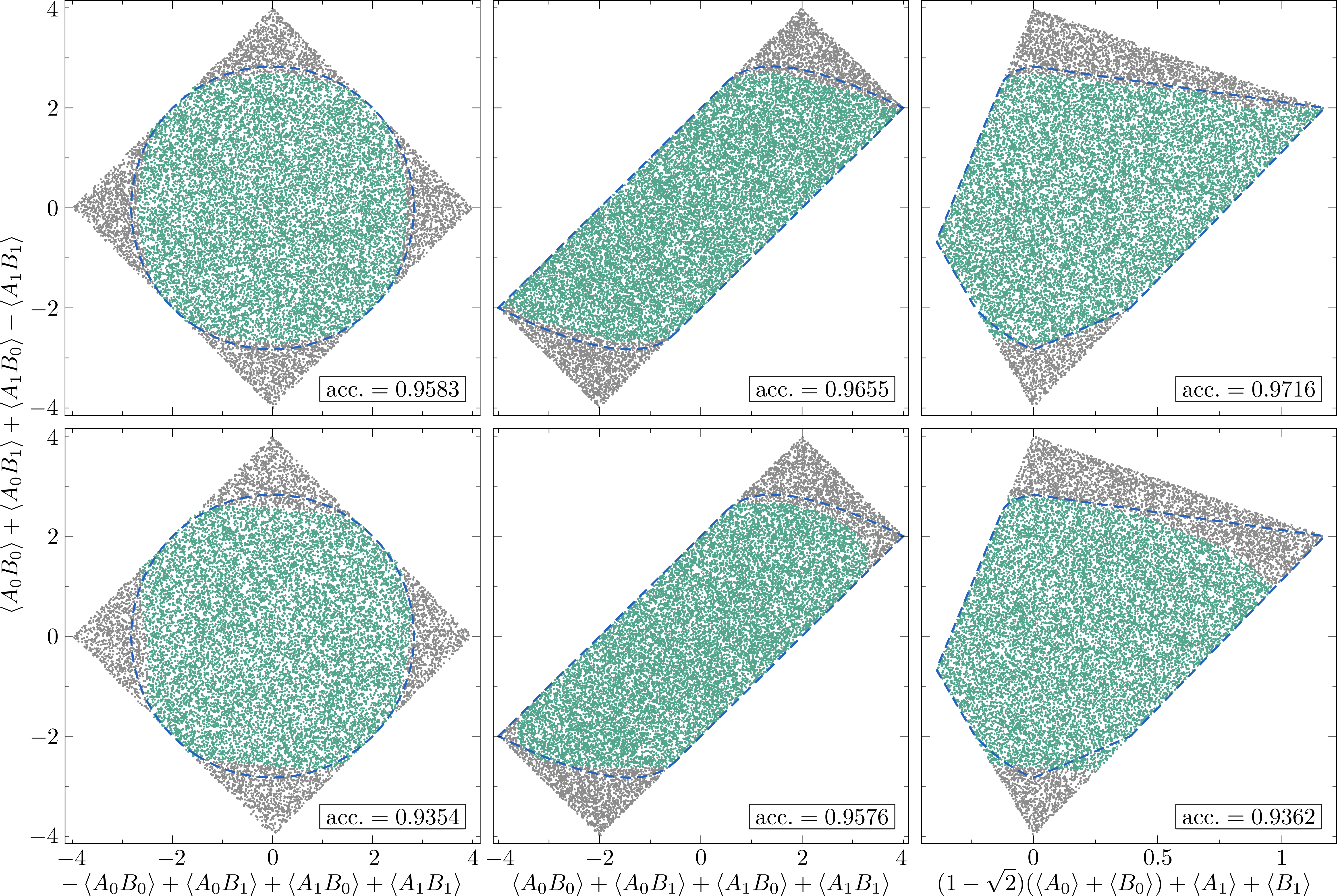}
    \caption{Two-dimensional slices of the entire CHSH scenario. The plots on the upper row correspond to the results for the SVM models trained with balanced data. The lower row depicts the results obtained for NNs using unbalanced data. The plots on the left and in the centre of the image correspond to the same correlation slices shown in FIGs.~\ref{fig: 4D SVM slices} and \ref{fig: 4D NN plots}. The right panels depict the slice introduced in FIG.~3 of Ref.~\cite{Goh18}, where we draw the boundary of the quantum set using level 3 of NPA.}
    \label{fig: 8D slices}
\end{figure*}

\begin{table*}[t!]
    \centering
    \setlength{\tabcolsep}{7pt}
    \caption{Accuracy measures for the SVM models in the entire CHSH space. The first column depicts the type of data used to train a given model, while the remaining columns present the values of the same accuracy measures used in the correlation models.}
    \small
    \begin{tabular}{ c | c | c | c | c | c }
        model & test acc. & train acc. & unbalanced acc. & spread acc. ($\sigma=10^{-2}$) & spread acc. ($\sigma=10^{-3}$)  \\ [0.5 ex] 
        \hline\hline
        unbalanced &  0.9527 & 0.9527 & --- & 0.5182 & 0.5013 \\ [0.15 ex]
        \hline
        balanced & 0.9610 & 0.9637 & 0.9153 & 0.5547 & 0.5009 \\ [0.15 ex]
        \hline
        offset ($\epsilon = 10^{-2}$) & 0.6417 & 0.7617 & 0.8266 & 0.6733 & 0.5157  \\ [0.15 ex]
        \hline
        offset ($\epsilon = 10^{-3}$) & 0.2893 & 0.6462 & 0.7959 & 0.7600 & 0.5335 \\ [0.15 ex]
        \hline
    \end{tabular}
    \label{table: SVM accuracies 8D}
\end{table*}

We then proceed with model development by generating samples containing $4\times 10^4$ points based on the previously used data types: unbalanced, balanced, and offset, with $\epsilon$ values of $10^{-2}$ and $10^{-3}$ for the offset data. Similar to the correlation space, here we also employed the RBF kernel of Scikit-learn and maintained a 70-15-15 ratio for training, testing, and validation sets, respectively. The resulting models can be divided into two cases. For the offset data, developing a model that consistently performed well across all accuracy measures proved challenging, as shown in TABLE~\ref{table: SVM accuracies 8D}. Unlike its correlation space analogues, the reduction of the offset value here did not enhance performance measures; instead, it caused more significant overfitting in the model with $\epsilon=10^{-3}$ compared to the one with $\epsilon=10^{-2}$. In contrast, increasing the feature dimension for the balanced and unbalanced models did not have a significant effect on the final model quality. Although their performance is still poor at points near the set boundary, as shown by the spread accuracy in TABLE~\ref{table: SVM accuracies 8D}, they show reasonable predictive capacity in other regions, as illustrated by the slices in the upper row of FIG.~\ref{fig: 8D slices}. The two first slices replicate the performance of the correlation models, which is extended to other regions of the representation space, as depicted in the rightmost plot.

\subsubsection{NNs on the entire CHSH space}

In the previous section, we have found that the training data sampled near the boundary is not well suited for neural network models. Consequently, for the 8-dimensional CHSH set, we consider only models trained with unbalanced data. We started by retraining the network we built for the benchmark on training sets of different sizes, starting with $10^4$ points. We first observe that the performance of the training model improves with the size of the dataset, up to a size of $\sim 5 \times 10^4$ points. While in terms of global accuracy the results of the training seem as good as those obtained in the 4-dimensional correlation space, with differences within $\sim 10^{-2}$, we find that the differences in accuracy on the slices studied above is about $5$ times larger. Interestingly, changes in the architecture of the network which increase the amount of resources like, for instance, doubling the amount of neurons in the input layer and/or adding extra hidden layers, do not seem to appreciably improve the performance of the trained models. It is not clear to us at the moment whether there is a combination of network architecture and hyperparameter values that may result in a better performing model, or if further improvement requires significantly larger training sets, which would represent an important practical limitation for these neural network models.

The global accuracy of the best model we could train in the conditions described above is $0.9967$, and in the lower row of figure of FIG.~\ref{fig: 8D slices} we show the performance of this model on the slices of the set previously introduced.  

\section{Discussion} \label{sec: discussion}

\noindent Using machine learning methods to solve problems in quantum physics is a relatively new phenomenon. Our goal was to conduct a comprehensive study of how these methods perform on a relatively simple and well-understood example. More specifically, we have chosen the problem of characterising the quantum set in the CHSH scenario, and we have explored various data science and machine learning models. It quickly became apparent that support vector machines and neural networks are the most promising candidates.

A key difference between the way ML is typically used and the way it is used in physics research is the source of data. In a research setting we are responsible for generating the data ourselves and one must be careful about how this is done exactly. In our case, uniform sampling leads to a highly imbalanced dataset. To fix this problem, we can either employ rejection sampling (which might be quite inefficient) or try to find a ``smart'' way of performing non-uniform sampling. However, one must be careful, because the latter might implicitly introduce some bias in the data resulting from our subjective opinion on what it means to be ``smart''.\footnote{Note that the same problem arises in the conventional setting where ML models are trained on pre-existing data, where the choice of how to clean and filter the data is also somewhat subjective. However, in our case, we are dealing with a particularly severe variant of this problem.}

Another observation that we made is that standard ML models are designed to perform well ``on average''. However, in physics research the usual focus is on analysing some specific phenomenon, e.g. determining the quantum value of a Bell functional or analysing a specific boundary region. Indeed, in research one almost exclusively cares about the ``hard'' cases, so average performance is not that relevant. For such tasks, the ML models do not seem to be a great fit.

Overall, both the SVMs and NNs exhibit reasonable performance, however, at some point adding more resources (e.g.~adding more data or increasing model complexity) leads to diminishing returns. The slices we plotted for the 4-dimensional set look quite good, but one can see some imperfections. For the slices of the 8-dimensional set, the imperfections are clearly visible. Note that the input dimension of 4 or 8 is very low for ML methods. Nevertheless, generating a data sample that is representative for the 8-dimensional correlation space is highly non-trivial.

Finally, we would like to make a comment on the approach of forcing the neural networks to be convex by restricting the coefficients to be negative. Convexity is not a concept that often appears in real-world data.\footnote{In fact, we are not aware of a single real-world dataset in which convexity plays a role.} Hence, we were excited to be able to apply this trick to data where convexity is indeed relevant. However, this did not lead to a visible improvement, which might suggest that combining ML approaches with concepts from abstract mathematics is not always fruitful. 

Overall, we have found that constructing an ML model that performs reasonably well on a relatively simple problem is easy. However, boosting its performance further is not that straightforward and at some point one has to start being smart about how the data is generated and how the model is chosen. At this point, however, there is a risk of injecting our own biases into the data without even realising it. Constructing models for larger problems is also challenging due to the sheer amount of data required to represent a high-dimensional space.

We believe that our findings will contribute to a better understanding of the types of problems where ML can provide an advantage over currently existing methods.

\section*{Acknowledgements}

\noindent We acknowledge support from the National Science Centre (NCN), Poland under the SONATA project ``Fundamental aspects of the quantum set of
correlations'' (grant No.~2019/35/D/ST2/02014). G.P.A. gratefully acknowledges Poland's high-performance Infrastructure PLGrid (HPC Centre: ACK Cyfronet AGH) for providing computer facilities and support (computational grant No. PLG/2023/016878). N.G. acknowledges support from CONICET and CIC (R.R.) of Argentina, CONICET PIP Grant No. 11220200101877CO

\bibliographystyle{apsrev4-1}
\bibliography{references}

\appendix

\vspace{20mm}

\section{\texorpdfstring{$\QC_{1+AB}$}{} \emph{vs}. \texorpdfstring{$\QC_{6 \times 6}$}{}} \label{app: Q1AB evaluation}

\noindent In this appendix, we leverage the approaches described in Secs.~\ref{subsec: sampling near the boundary} and \hyperref[subsec: steered see-saw]{C} to numerically compare the level $1+AB$ of NPA with the set of quantum realisations with local dimension $6$, referred to as $\QC_{1+AB}$ and $\QC_{6 \times 6}$, respectively. The choice of a local dimension of 6 is based on its computational feasibility within the steered see-saw algorithm and the fact that, for $d \ge 4$, $\LC \in \QC_{d \times d}$ \cite{Donohue15}. This allows us to limit our focus to directions pointing towards one of the CHSH facets, since the boundaries of the two analysed sets coincide in other directions.

\begin{figure}[b!]
    \centering
    \includegraphics[width=86mm]{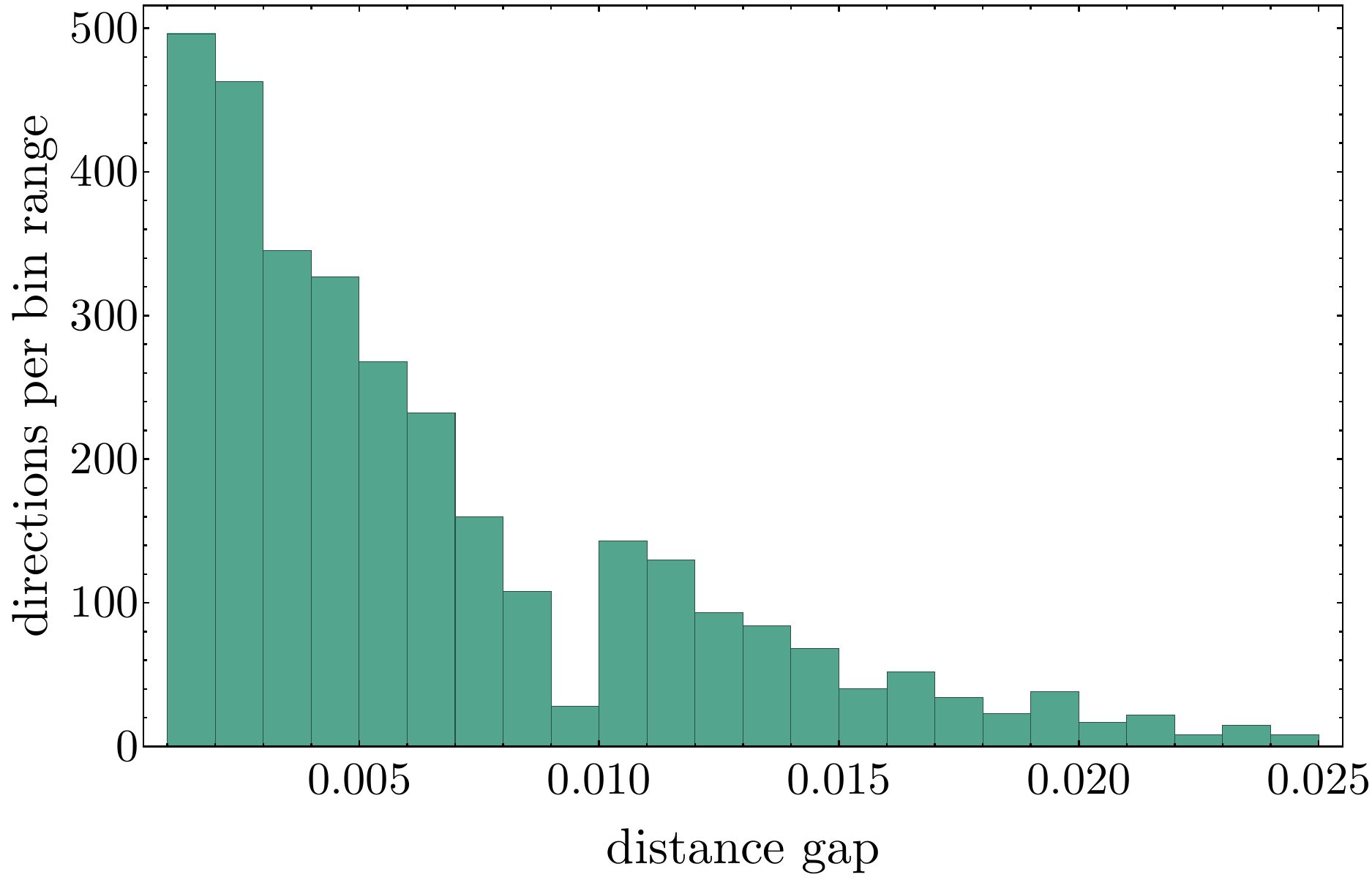}
    \caption{The distribution of distances for the directions where the result was greater than $10^{-3}$. Each bin has a length of also $10^{-3}$. In the optimisation presented in Eq.~\eqref{eq: optimise rho}, the minimised quantity is the distance between the variable encoded in the correlation $\mathbf{q}$ and the target point $\mathbf{p}_b = \lambda^* \mathbf{u}$, with the direction $\mathbf{u}$ not included in the problem constraints. Consequently, the interior estimate of $\mathbf{p}_b$ does not necessarily align with $\mathbf{u}$, though the angle between the two points relative to the polytope’s centre decreases proportionally with the gap between them.}
    \label{fig: histogram}
\end{figure}

We then begin with a sample of $10^4$ directions obtained by intersecting random directions with the local facet defined in Eq.~\eqref{eq: classic CHSH}. For each of the selected directions, we compute a pair of points: one on the boundary of $\QC_{1+AB}$, using the approach outlined in Eq.~\eqref{eq: NPA SDP}, and an interior estimate within $\QC_{6 \times 6}$, using the steered see-saw with 50 random initializations or until the problem value is lower than $10^{-7}$. The distances between each corresponding pair are then calculated and analysed. Among the total directions considered, 9220 showed a distance of less than $10^{-2}$, with 6793 of these falling below $10^{-3}$. For 5903 directions, the gap was smaller than the initially defined threshold of $10^{-7}$. The plot in FIG.~\ref{fig: histogram} illustrates the distribution of distances for the pairs that are at least $10^{-3}$ apart. As a result, considering that the simplex comprised by the facet in Eq.~\eqref{eq: classic CHSH} covers only 0.21\% of the total solid angle measured at $\mathbb{R}^8$ from its origin, we observe that the method described in Sec.~\ref{subsec: sampling near the boundary} correctly classifies points for about 99.87\% of the total directions when $\epsilon = 10^{-2}$ and 99.46\% when $\epsilon = 10^{-3}$.

It is worth noting that, since steered see-saw does not guarantee obtaining a point on the boundary of $\QC_{6 \times 6}$, the quality of the estimate depends on the number of random initialisations used. For this reason, in an attempt to obtain a more realistic number for the quantity of seeds required, we carried out the test that can be summarised by the data in TABLE~\ref{table: local dim volumes}. In summary, we produce a sample of again $10^4$ points uniformly distributed in the simplex defined by the local facet of Eq.~\eqref{eq: classic CHSH} and the PR box corresponding to this facet. Then, we eliminate points that did not belong to the superset $\QC_4$ of NPA and classified the remaining points with steered see-saw using initially 5 seeds and a given local dimension. As before, if the problem value obtained is smaller than the threshold of $10^{-7}$ for a given dimension, the evaluated point is considered as belonging to $\QC_{d \times d}$. Initially, our idea was to compute the volume resulting from the intersection of the simplex with the sets $\QC_{d \times d}$, for $d \le 16$. However, this quickly proved to be more challenging than anticipated. As the dimension increased throughout the computation rounds, the process became unfeasible for $d = 10$. Nonetheless, during the following rounds, by decreasing the local dimension and increasing the number of seeds, we were able to classify most of the remaining points.

The data presented allows for two general conclusions. Firstly, the number of seeds used in the computation described at the beginning of this appendix is suboptimal, indicating that the superset $\QC_{1+AB}$ is an even better approximation of $\QC$ for developing an ML classification model. Secondly, numerically distinguishing between sets of quantum realisations with fixed local dimension for $d \ge 10$ is difficult. This is because it is not only computationally expensive but also due to the small volume difference between these sets in higher dimensions.

\begin{table}[b!]
    \centering
    \setlength{\tabcolsep}{5.5pt}
    \caption{Classification of a sample of $10^4$ points uniformly distributed in the simplex comprised over the local facet defined in Eq.~\eqref{eq: classic CHSH} and the box $\mathbf{p}_\text{PR} = [0,\, 0,\, 0,\, 0,\, 1,\, 1,$ $ 1, -1]$. After removing points not belonging to the level 4 of NPA, the remaining 8563 points were classified in 17 rounds, varying the local dimension and the number of seeds. The second column shows the local dimension used, the third column shows the number of random initialisations, and the fourth column shows the number of points for which the quantum realisation found had a distance less than $10^{-7}$ from the target point. The last column indicates the total computation time per round, formatted as days:hours:mins. Note that in the 8th round, with $d=9$, the computation time increased such that further increases in dimension became prohibitive. In subsequent rounds, the local dimension was reduced to $d=6$ and $8$, and the number of seeds was increased. The total CPU time was 158 days, 1 hour, and 26 minutes, with only 54 points remaining unclassified.}
    \small
    \begin{tabular}{ c || c | c | c | c }
        round \# & local $d$ & No. seeds &  No. $\in \QC_{d \times d}$ & CPU time \\ [0.5 ex] 
        \hline\hline
        1 &  2 & 5 & 2271 & 01:07:30 \\ [0.15 ex]
        \hline
        2 &  3 & 5 & 2597 & 02:22:38 \\ [0.15 ex]
        \hline
        3 &  4 & 5 & 2097 & 01:07:45 \\ [0.15 ex]
        \hline
        4 &  5 & 5 & 496 & 01:12:32 \\ [0.15 ex]
        \hline
        5 &  6 & 5 & 390 & 03:10:23 \\ [0.15 ex]
        \hline
        6 &  7 & 5 & 171 & 04:12:04 \\ [0.15 ex]
        \hline
        7 &  8 & 5 & 94 & 06:07:27 \\ [0.15 ex]
        \hline
        8 &  9 & 5 & 56 & 17:11:05 \\ [0.15 ex]
        \hline
        9 &  6 & 50 & 158 & 11:03:25 \\ [0.15 ex]
        \hline
        10 &  8 & 50 & 37 & 16:20:39 \\ [0.15 ex]
        \hline
        11 &  6 & 100 & 77 & 12:11:57 \\ [0.15 ex]
        \hline
        12 &  6 & 150 & 24 & 11:04:53 \\ [0.15 ex]
        \hline
        13 &  6 & 200 & 14 & 11:05:17 \\ [0.15 ex]
        \hline
        14 &  6 & 250 & 9 & 12:06:02 \\ [0.15 ex]
        \hline
        15 &  6 & 300 & 9 & 14:20:05 \\ [0.15 ex]
        \hline
        16 &  6 & 350 & 7 & 14:04:17 \\ [0.15 ex]
        \hline
        17 &  6 & 400 & 2 & 15:01:20 \\ [0.15 ex]
        \hline
    \end{tabular}
    \label{table: local dim volumes}
\end{table}

\end{document}